\documentclass[%
 reprint,
superscriptaddress,
 amsmath,amssymb,
 aps,
prb,
floatfix,
]{revtex4-2}

\usepackage[T1]{fontenc} 
\usepackage[utf8]{inputenc} 
\usepackage{lmodern} 
\usepackage{amsmath} 
\usepackage{amssymb} 
\usepackage{siunitx}
\usepackage{physics}
\usepackage{bm} 
\usepackage{graphicx}
\usepackage{xcolor,soul}
\usepackage{hyperref}

\begin{document}
\title{Integrated molecular optomechanics with hybrid dielectric-metallic resonators }
\author{Ilan Shlesinger}
\affiliation{Center for Nanophotonics, AMOLF, Science Park 104, 1098 XG Amsterdam, The Netherlands}
\author{K\'{e}vin G. Cogn\'{e}e}
\affiliation{Center for Nanophotonics, AMOLF, Science Park 104, 1098 XG Amsterdam, The Netherlands}
\affiliation{LP2N, Institut d'Optique Graduate School, CNRS, Univ. Bordeaux, 33400 Talence, France}
\author{Ewold Verhagen}
\affiliation{Center for Nanophotonics, AMOLF, Science Park 104, 1098 XG Amsterdam, The Netherlands}
\author{A. Femius Koenderink}
\affiliation{Center for Nanophotonics, AMOLF, Science Park 104, 1098 XG Amsterdam, The Netherlands}


\begin{abstract}
Molecular optomechanics stems from the description of Raman scattering in the presence of an optical resonator using a cavity optomechanics formalism. We extend the molecular optomechanics formalism to the case of hybrid dielectric-plasmonic resonators, with multiple optical resonances and with both free-space and waveguide addressing.
We demonstrate how the Raman enhancement is the product of a pump enhancement and a modified LDOS, that simply depend on the complex response functions of the hybrid system. The Fano lineshapes that result from hybridization of a broadband and narrowband modes allows reaching strong Raman enhancement with high-Q resonances, paving the way towards sideband resolved molecular optomechanics. The model allows prediction of the Raman emission ratio into different output ports and enables demonstrating a fully integrated high-Q Raman resonator exploiting multiple cavity modes coupled to the same waveguide.

\end{abstract}
\maketitle
\section{Introduction}



It is well understood that surface enhanced Raman spectroscopy (SERS)~\cite{Fleischmann1974,Jeanmaire1971,Albrecht1977} benefits both of the electromagnetic field enhancement of the pump field driving the Raman process, and the plasmonically generated local density of states (LDOS) enhancement for the emission at the Stokes and anti-Stokes sidebands~\cite{LeRu2009PrinciplesSpectroscopy,chu_2010,Ye2012}.
Recently, the new formalism of molecular optomechanics was introduced, showing the analogy of SERS with cavity optomechanics~\cite{Roelli2016MolecularScattering}. It describes the Raman process as an optomechanical interaction between a localized plasmonic resonance and a molecule's nuclear motion. The coupling of light and motion stems from a dispersive shift of the plasmonic resonance upon the molecule's vibrational displacement~ \cite{Roelli2016MolecularScattering,KamandarDezfouli2017QuantumResonators,Schmidt2016QuantumCavities,Schmidt2017LinkingSERS}. The cavity optomechanics viewpoint allows a consistent description of optical forces on the molecule's vibration inducing quantum and dynamical backaction~\cite{Schmidt2017LinkingSERS}, which were previously described phenomenologically as vibrational pumping and as plasmonic asymmetry factor~\cite{Maher2008}.
Furthermore, with the correct description of coherence in the optomechanical interaction, new phenomena such as collective effects have been predicted~\cite{zhang_2020}, and within state of the art in plasmonic nano- and picocavities~\cite{Benz2016Single-moleculepicocavities} one could envision promising applications such as
coherent quantum state transfer and entanglement between photons and phonons~\cite{Palomaki2013EntanglingFields,Palomaki2013CoherentOscillator,Reed2017FaithfulMotion,Roelli2020} at high frequencies (1-100 THz), where no cooling is required.

Cavity-optomechanics  often operates in the so-called `resolved sideband regime', wherein the mechanical frequency exceeds the optical linewidth~\cite{Aspelmeyer2014CavityOptomechanics}. This, for instance, is deemed crucial for cooling of macroscopic motion through selectively enhancing anti-Stokes scattering~\cite{Cohadon1999}, and to reach coherent conversion from photons to phonons and back~\cite{Palomaki2013EntanglingFields,Palomaki2013CoherentOscillator,Reed2017FaithfulMotion}. The molecular optomechanics equivalence would be to have access to optical resonators with linewidths narrower than the vibrational frequency of the molecular species at hand, yet nonetheless exceptionally good confinement of the electric field for large coupling to the Raman dipole. This regime is not easily reached with plasmonics, as resonators typically have quality factors $Q\sim 20$, meaning linewidths larger than or comparable to vibrational frequencies (500-1500 cm$^{-1}$). Conversely,  conventional higher $Q$ dielectric resonators typically have poor mode confinement, and hence poor SERS enhancement~\cite{Giannini2011}. In the last few years hybrid photonic-plasmonic resonators have emerged in which hybrid resonances of dielectric microcavities coupled to plasmonic antennas are used~\cite{Foreman2013LevelBiodetection,Thakkar2017SculptingHybridization,Pan2020,Frimmer2012SuperemittersLimit,KamandarDezfouli2017ModalSystem,Barth2010NanoassembledCoupling,Gurlek2018ManipulationStrong-Coupling}. Theoretical and experimental evidences points at plasmonic confinement ($<\lambda^3/10^5$) with microcavity quality factors ($Q > 10^3$)~\cite{Palstra2019HybridCryostat,Doeleman2020}. 

In this work we report on a semi-classical molecular optomechanics model for waveguide-addressable multi-resonant hybrid photonic-plasmonic resonators coupled to molecular mechanical oscillators. This work has several important novelties.  First, in evaluating the SERS enhancement, previous work has generally approximated the optical system as a single Lorentzian resonator~\cite{Dezfouli2019MolecularModes}. In contrast, even the simplest  hybrid resonators  show Fano-lineshapes in their response function~\cite{Ameling2013}, responsible for the SERS enhancement factors. Thus we expect SERS in hybrids to be controlled by a spectrally complex structure in LDOS, encompassing high-$Q$ Fano lines, and a low-$Q$ plasmon-antenna like contribution.   Secondly, we extend this work from simple hybrid dielectric-photonic resonators to hybrids in which a single antenna hybridizes with multiple microcavity modes. This allows further control of SERS, through the accurate engineering of the structured photonic reservoir for Stokes, pump, and anti-Stokes frequencies independently. This scenario could be achieved with any whispering gallery mode (WGM) cavity system, with free spectral ranges that match vibrational frequencies~\cite{Soltani2018StimulatedResonators,Liu2021}. Finally, a main generalization of our work over earlier works is that we include input-output channels. Indeed, in prospective  SERS experiments with hybrid dielectric photonic resonators, a waveguide can be specifically and efficiently interfaced with the cavity, to address hybrid resonances~\cite{Peyskens2016SurfacePlatform}.
Using different input and output channels opens up new scenarios for detection schemes, like, for instance, pumping from free-space and collecting the Raman scattered power distributed over one or different output waveguide ports.
This means it is important to determine the ideal pumping and collection scheme.  Our semi-analytical model illustrates the potential and tradeoffs for waveguide-addressable hybrid photonic-plasmonic resonators for physically relevant parameters for cavities and plasmon antennas taken from full wave numerical modelling~\cite{Doeleman2016Antenna-CavityLinewidth}. We derive realistic and quantitative predictions for SERS enhancements that can be compared with those obtained with the usual bare plasmon nanoparticle antennas.  

\begin{figure}[h]
\centering
 \includegraphics[width=0.42\textwidth]{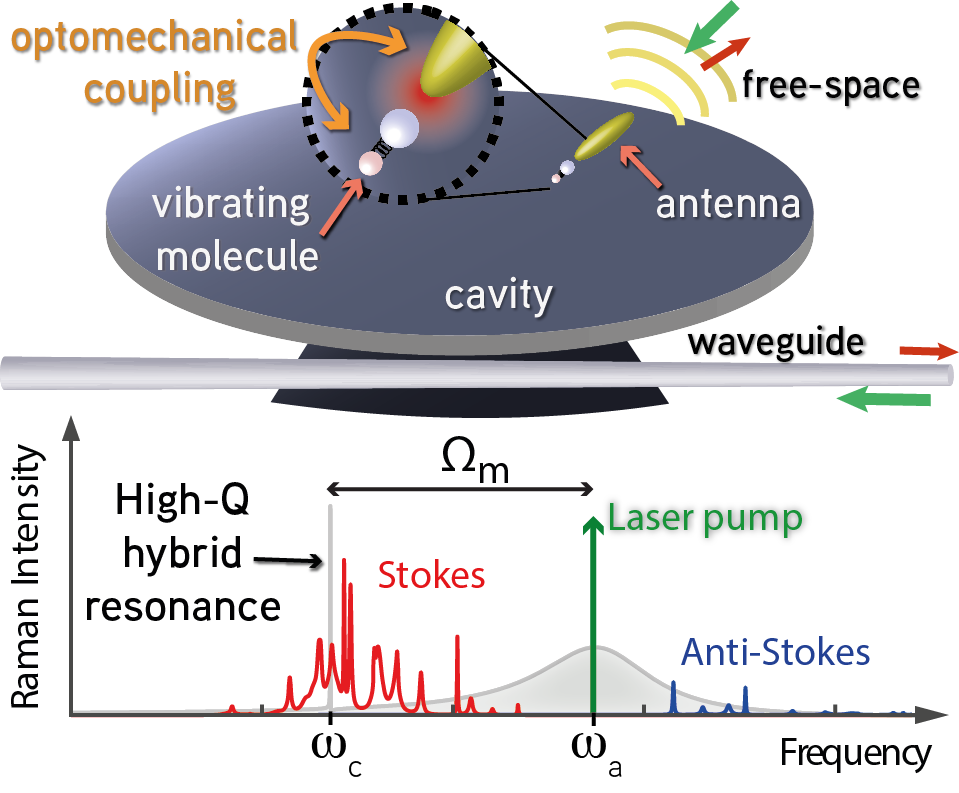}
\caption{Raman scattering enhanced by a hybrid dielectric-plasmonic resonator. Top: sketch of a typical system: the spectrally narrow modes of a dielectric cavity hybridize with a plasmonic antenna resulting in high-Q small mode volume resonances, ideal for sideband resolved molecular optomechanics. Light can couple in and out through different ports such as the free-space or waveguides. Bottom: the hybrid system can be used to enhance both the laser pump and Raman sidebands, even in the sideband resolved regime, where the linewidth of the optical resonances are narrower than the mechanical frequency $\Omega_m$.
} 
\label{fig:Skecth_molecularOM}
\end{figure}

\section{Hybrid molecular optomechanics formalism}
We first consider a single mode hybrid resonator composed of a plasmonic antenna coupled to a high-Q dielectric cavity. The model is based on semi-classical Langevin equations~\cite{Roelli2016MolecularScattering,Schmidt2017LinkingSERS} where a plasmonic antenna, described as a polarizable electrodynamic dipole scatterer, is coupled to a microcavity mode, quantified by a resonance frequency, mode volume, and intrinsic damping rate. In short, this model can be reduced to a description in terms of coupled equations of motion for two harmonic oscillators~\cite{Doeleman2016Antenna-CavityLinewidth}. The single cavity mode is described by the field amplitude $c(\omega)$, such that $\abs{c(\omega)}^2$ is the normalized energy contained in the mode, with a resonance frequency $\omega_c$ and a damping rate $\kappa$. The excitation of the antenna is quantified by its induced dipole moment $p$, which derives from its polarizable nature. We assume a polarizability with resonance frequency $\omega_a$,  oscillator strength $\beta$, and a total damping rate $\gamma_a(\omega)=\gamma_i+\gamma_{\mathrm{rad}}(\omega)$, taking into account intrinsic ohmic losses and frequency-dependent radiation losses assumed in vacuum $\gamma_{\mathrm{rad}}(\omega)=\frac{\beta \omega^2}{6\pi\epsilon_0 c^3}$~\cite{Novotny2006PrinciplesNano-Optics}. The dynamic antenna polarizability is then given by $\alpha_0(\omega)=\frac{\beta}{\omega_a^2-\omega^2-i\omega\gamma_a}$.
Similarly to the cavity mode mode, the antenna field will be described  by the field amplitude $a(\omega)=\frac{\omega}{\sqrt{2\beta}}p$.
We consider each of the two optical resonators to be coupled to a unique port: a waveguide for the cavity mode and free-space for the antenna.
A vibrating molecule is placed in the hotspot of the antenna $\mathbf{r}_0$, and its vibration, corresponding to the stretching or compression of a specific molecular bond, is also described as an harmonic oscillator with a mechanical coordinate $x_m$ a resonance frequency $\Omega_m$ and a decay rate $\Gamma_m$. 
The parametric Raman process is described as an optomechanical interaction between the molecule's vibration and the optical fields at its position~\cite{Roelli2016MolecularScattering,Schmidt2017LinkingSERS}.
The Langevin equations for the three harmonic oscillators (antenna, cavity and mechanical mode), are written in the rotating frame of a laser pump of frequency $\omega_L$ (see appendix~\ref{sec:Sup1}):
\begin{widetext}
\begin{align}\label{eq:langevin}
    \dot{a}+(-i(\omega_L-\omega_a)+\gamma_a/2)a - i x_m (G_{a} a + G_{\text{cross}} c) -i Jc &=\sqrt{\eta_{\text{in},a}\gamma_{\mathrm{rad}}}\,s_{\text{in},a}\nonumber
    \\
    \dot{c}+(-i(\omega_L-\omega_c)+\kappa/2)c - i x_m (G_{c} c + G_{\text{cross}} a) -i Ja &=\sqrt{\eta_{\text{in},c}\kappa}\,s_{\text{in},c}\nonumber
    \\
    \ddot{x}_m+\Omega_m^2x_m +\Gamma_m\dot{x}_m-\frac{\hbar}{m}(G_{a}|a|^2+G_{c}|c|^2+G_{\text{cross}}(a^*c+ac^*))
    &=F_\text{ext}/m\,.
\end{align}
\end{widetext}
In these equations the cavity and antenna are linearly coupled through a hybridization strength $|J|$. This term describes the purely electromagnetic coupling between the two resonators: the antenna driving the cavity mode or the cavity polarizing the antenna. The magnitude of $J$ depends on the confinement of the cavity field at the antenna position parametrized by the mode volume $V_c$, the oscillator strength of the antenna $\beta$, as well as the orientation of the antenna with respect to the cavity field polarization. It is written as $|J|=\frac{1}{2} \sqrt{\frac{\beta}{\epsilon_0 V_c}}$, where $V_c$ is the effective cavity mode volume felt by the antenna $V_{c}=\frac{2}{\epsilon_0  \abs{\tilde{E}_{c}(\mathbf{r_0})}^2}$, with  $\tilde{E}_{c}(\mathbf{r_0})=\mathbf{e_p}\cdot\mathbf{\tilde{E}}_{c}(\mathbf{r_0})$ the normalized mode profile of the cavity field along the antenna polarization axis $\mathbf{e_p}$ at the position of the antenna, and simply written as $\tilde{E}_{c}$ in the following. The hybrid coupling then appears as a dipolar coupling rate between the antenna dipole and the cavity field. 

The optomechanical coupling arises from a modification of the antenna and the cavity resonance frequencies due to the vibration of the molecule and is described by the optomechanical coupling strengths $G_{a}$ and $G_{c}$ between the molecule with either the antenna or the cavity mode, and can be evaluated using first-order perturbation theory yielding~\cite{Roelli2016MolecularScattering}:
\begin{equation}\label{eq:def_G}
    G_{a,c}=\frac{\omega_{a,c}}{2\epsilon_0\epsilon V_{a,c}}\frac{\partial \alpha_m}{\partial x_m}=\frac{\omega_{a,c}}{4}\tilde{E}_{a,c}(\mathbf{r_0})\frac{\partial \alpha_m}{\partial x_m}\,,
\end{equation}
where, similarly to the cavity mode, we have introduced an antenna effective mode volume $V_{a}=\frac{2}{\epsilon_0 \epsilon \abs{\tilde{E}_{a,c}(\mathbf{r_0})}^2}$, with $\tilde{E}_{a}(\mathbf{r_0})$ the normalized mode profile of the antenna, evaluated at the position of the molecule, and simply written $\tilde{E}_{a}$ in the following.
The overlapping optical fields of the antenna and the cavity at the position of the vibrating molecule result in a crossed optomechanical coupling whose coupling strength can be approximated as $G_{\text{cross}}\simeq\sqrt{G_{a}G_{c}}$ (see Appendix~\ref{sec:Sup1}).
Two different inputs are considered for the laser pump, either a free-space input where a far-field pump directly polarizes the antenna, or, waveguide input selectively exciting the cavity modes, with input amplitude and coupling efficiencies $s_{\text{in},a}$ and $\eta_{a,in}$ for the antenna, and $s_{\text{in},c}$ and $\eta_{in,c}$ for the cavity mode. The input amplitudes are normalized such that $\abs{s_\text{in}}^2$ is the optical power entering at a given port.
$F_\text{ext}$ describes the input mechanical fields, which is here considered to be only thermal fluctuations.
We thus arrive to the same coupled equations as derived in~\cite{Roelli2016MolecularScattering}, however extended to take into account multiple optical modes and different inputs. We note that while the equations and phenomena considered here are classical, they could readily be extended to include quantum fluctuations by introducing noise terms with appropriate correlators.

In the present work, we are only interested in the low-cooperativity regime, which is the most experimentally relevant~\cite{Roelli2016MolecularScattering}, and we can thus neglect the back-action of the optical fields on the mechanical resonance, i.e. consider only thermal mechanical fluctuations $x_m$ due to $F_\text{ext}$. The optical resonator amplitudes in the third equation of Eqs~\ref{eq:langevin} will then be neglected, which discards the laser quantum back action as well as dynamical back action on the mechanical mode. 
The noise spectral density of the molecule's vibration is then given by the quantum Nyquist formula~\cite{Barker1972}:
\begin{align}\label{eq:S_xx}
S_{xx}(\Omega)=x_{\mathrm{zpf}}\Gamma_m\bigg[\frac{\bar{n}_{\mathrm{th}}}{(\Omega-\Omega_m)^2+(\Gamma_m/2)^2}+ \nonumber\\
\frac{\bar{n}_{\mathrm{th}}+1}{(\Omega+\Omega_m)^2+(\Gamma_m/2)^2}\bigg] ,
\end{align}
with the mean phonon occupation $n_{\mathrm{th}}=(\mathrm{exp}(\frac{\hbar\Omega_m}{k_b T}-1)^{-1}$ for a bath at temperature $T$, and the zero point amplitude $x_{\mathrm{zpf}}$.
These mechanical fluctuations will translate into optical Raman signal through the optomechanical coupling with the antenna and cavity modes as described by the two remaining Langevin equations for the optical fields. These can be linearized by decomposing the fields in a steady-state plus a fluctuating part, $ a \rightarrow \bar{\alpha}_a +  a$, $c \rightarrow \bar{\alpha}_c +  c $ and $ x_m \rightarrow \bar{x}_m +  x_m$.  Finally the small frequency shift due to the steady-state mechanical displacement $\bar{x}_m\sim 0$ is absorbed in the definition of $\omega_a$ and $\omega_c$.
The solutions for the steady-state solutions our found by setting $a=c=0$ and we get:
\begin{align}
\begin{cases}
\bar{\alpha}_a=\chi_a'\left(\sqrt{\eta_{\text{in},a}\gamma_{\mathrm{rad}}}\,s_{\text{in},a}+iJ^*\chi_c \sqrt{\eta_{\text{in},c}\kappa}\,s_{\text{in},c}\right)\\
\bar{\alpha}_c=\chi_c'\left(\sqrt{\eta_{\text{in},c}\kappa}\,s_{\text{in},c}+iJ\chi_a \sqrt{\eta_{\text{in},a}\gamma_{\mathrm{rad}}}\,s_{\text{in},a}\right).
\end{cases}\label{eq:avgsol}
\end{align}
They correspond to the solution of a Rayleigh scattering process.
The susceptibilities of the bare cavity mode $\chi_c$ and antenna mode $\chi_a$ are
\begin{align}
\begin{cases}
    \chi_a(\omega)=\frac{i}{\omega-\omega_a+i\frac{\gamma_a}{2}}=-i\frac{2\omega}{\beta}\alpha_0(\omega)\\
    \chi_c(\omega)=\frac{i}{\omega-\omega_c+i\frac{\kappa}{2}}
\end{cases}.
\end{align}
The antenna and cavity response are modified by the hybrid coupling $J$  which yields new hybridized susceptibilities $\chi_c'$ and $\chi_a'$ for the two optical modes:
\begin{equation}\label{eq:suscept-hybrid}
     \chi_{a,c}'(\omega)=\frac{\chi_{a,c}}{1+\abs{J}^2\chi_a\chi_c}.
\end{equation}
They can be seen as the bare constituents susceptibilities, dressed by an infinite series of antenna-cavity scattering events. The antenna having a very broad response compared to the cavity, the hybridized susceptibility $\chi_a'$ will display a Fano resonance at the frequency of the cavity~\cite{Limonov2017}.

The fluctuating part of the field ($\Omega\neq0$), responsible for the Raman scattering, is expressed in the frequency domain, and we keep only terms that are first order in the fluctuations (i.e. $x_m a $, $x_m c\rightarrow0$):
\begin{equation}
\begin{cases}
\left(\omega_L+\Omega-\omega_a+i\frac{\gamma_a}{2}\right)a+ Jc = - x_m (G_{a} \bar{\alpha}_a + G_{\text{cross}} \bar{\alpha}_c)\\
\left(\omega_L+\Omega-\omega_c+i\frac{\kappa}{2}\right)c+Ja = -x_m (G_{c} \bar{\alpha}_c + G_{\text{cross}} \bar{\alpha}_a).
\label{eq:hybridfluctuation}
\end{cases}
\end{equation}
Note that the fluctuations are evaluated at the frequency $\omega_L+\Omega$. To evaluate the stokes or anti-Stokes sidebands we will later set $\Omega=\mp\Omega_m$.
The right-hand side of the equations show that the source terms for the optical fluctuations arise from a sum of direct and crossed optomechanical coupling with the mechanical vibration. The first is given with a rate $G_a$ or $G_c$, and the second through crossed optomechanical coupling $G_\text{cross}$, and are directly proportional to the steady state solutions obtained previously. Both of these processes are described by an effective optomechanical coupling $G_{a,c}^\text{eff}=G_{a,c} \bar{\alpha}_{a,c} + G_{\text{cross}} \bar{\alpha}_{c,a}$ taking both the coupling rate and the steady state solutions into account. We finally obtain the solutions for the fluctuations:
\begin{align}
\begin{cases}
a(\omega_L+\Omega)=i\chi_a' \left(G_{a}^\text{eff}+iJ^*\chi_c G_{c}^\text{eff}\right)x_m(\Omega)\\
c(\omega_L+\Omega)=i\chi_c' \left(G_{c}^\text{eff}+iJ^*\chi_a G_{a}^\text{eff}\right)x_m(\Omega)
\label{eq:solu_with_G_eff}
\end{cases},
\end{align}
with all the suceptibilities evaluated at the emission frequency $\omega_L+\Omega$.
The antenna and cavity fluctuations $a$ and $c$ appear as a transduction of mechanical fluctuations $x_m$ with a modified response due to the hybrid coupling characterized by $J$. 
The case with only a bare antenna corresponding to the usual SERS experiments is retrieved by setting $J=G_\mathrm{cross}=0$, yielding:
\begin{equation}\label{eq:a_bare}
    a^{\mathrm{bare}}(\omega_L+\Omega)=ix_m(\Omega)\chi_a(\omega_L)  G_{a}\bar{\alpha}_a \,.
\end{equation}
Following the optomechanical formalism put forward by~\cite{Roelli2016MolecularScattering}, with no backaction on the mechanical mode, we have arrived at a set of coupled classical equations where the optical fluctuations of the modes (inelastic process $\omega\neq\omega_L$) are driven by mechanical vibrations of the molecule. 

Raman spectra scattered by the antenna to the far-field $S_{\text{ant}}$ and the cavity to the waveguide $S_{\text{cav}}$ can be immediately expressed as
\begin{equation}
\begin{cases}
    S_{\mathrm{ant}}(\omega_D=\omega_L+\Omega)=\eta_\mathrm{a,out}\gamma_{\mathrm{rad}}\abs{a(\omega_D)}^2
    \\
    S_{\mathrm{cav}}(\omega_D=\omega_L+\Omega)=\eta_\mathrm{c,out}\kappa\abs{c(\omega_D)}^2
\end{cases}\label{eq:spec_out_hyb}
\end{equation}
where $\omega_D$ is the frequency of detection.

To obtain Raman enhancements factors, the spectra are normalized by the emission of the molecule in the homogeneous medium, in the absence of a resonator, given for the same excitation and collection conditions. This reference situation is modelled as the scattering of the Raman dipole of the molecule~\cite{LeRu2009PrinciplesSpectroscopy}, in which
\begin{equation}
    p_R(\omega_D=\omega_L+\Omega)=\frac{\partial \alpha}{\partial x_m}x_m(\Omega)E_{\mathrm{inc}}(\omega_L)
\end{equation}
where $E_\mathrm{inc}$ is the incident field at the position of the molecule (see Appendix~\ref{sec:Sup2}).
Using Larmor's formula~\cite{Jackson1999} one obtains the reference Raman scattered spectrum for the molecule in a homogeneous medium of index $n=1$:
\begin{equation}
    S_\mathrm{ref}(\omega_D,\omega_L)=\frac{\omega_D^4}{12\pi \epsilon_0 c^3}\abs{\frac{\partial \alpha}{\partial x_m}}^2 S_{xx}(\Omega)\abs{E_\mathrm{inc}(\omega_L)}^2 ,
    \label{eq:S_ref}
\end{equation}
where we have replaced $\abs{x_m}^2\rightarrow S_{xx}$.

By replacing $a$ and $c$ by their expression of Eq.~\ref{eq:solu_with_G_eff}, and using Eq.~\ref{eq:def_G} one can write the Raman spectrum of the antenna and the cavity as the product of three terms:
\begin{align}
S_\mathrm{ant,cav}=\text{Pump enh.}\times\text{LDOSC}_\text{ant,cav}\times  S_\text{ref} \,  ,
\label{eq:factor_2}
\end{align}
i.e., the reference spectrum $S_\text{ref}$ enhanced both by a pump enhancement term and a collected LDOS (LDOSC) in either output port. The pump enhancement is given by
\begin{equation}\label{eq:pump_enh}
    \text{Pump enh.}=\left|\frac{\bar{\alpha}_a\tilde{E}_a+\bar{\alpha}_c\tilde{E}_c}{E_{\mathrm{inc}}}\right|^2,
\end{equation}
and corresponds to the field enhancement due to the optical hotspots compared to the incident field. The total field at the molecule's position (neglecting the incident field direct contribution) shows a coherent mixing of cavity and antenna contributions.
The Raman emission is also enhanced by the collected LDOS, which, depending on the assumed collection channel, i.e.,  through the free-space or the waveguide port, reads 
\begin{align}\label{eq:LDOSac}
    \text{LDOSC}_\text{ant}(\omega)&=\eta_{\mathrm{a,out}}\gamma_\mathrm{rad}\frac{3\pi\epsilon_0 c^3}{2\omega^2} \times \nonumber \\
    & \abs{\chi_a'(\omega) \left(\tilde{E}_a^*+iJ^*\chi_c(\omega)\tilde{E}_c^*\right)}^2 , \nonumber \\
    \text{LDOSC}_\text{cav}(\omega)&=\eta_{\mathrm{c,out}}\kappa\frac{3\pi\epsilon_0 c^3}{2\omega^2} \times \nonumber\\
    &\abs{\chi_c'(\omega)\left(\tilde{E}_c^*+iJ^*\chi_a(\omega)\tilde{E}_a^*\right)}^2\, .
\end{align}
The total LDOS obtained by summing these two expressions can be cast as~\footnote{with an equality only if $\eta_{out}=1$, otherwise the intrinsic losses need to be added.}:
\begin{align}
    \text{LDOS}_\text{tot}=\frac{3\pi\epsilon_0 c^3}{\omega^2}\Im{\sum_{j=a,c}i\chi_j'|\tilde{E}_j|^2-2J\chi_a'\chi_c\tilde{E}_a\tilde{E}_c^*}
\end{align}
given by the sum of the LDOS of the hybridized antenna and cavity modes, along with a term arising from coherent interaction between the two resonators.

Both LDOSC expressions of Eqs.~\ref{eq:LDOSac} show a coherent coupling between antenna and cavity characterized by the effective susceptibilities \begin{align}
    \chi_{a,c}^\text{eff}= \left(\tilde{E}_{a,c}^*+iJ^*\chi_{c,a}\tilde{E}_{c,a}^*\right)\chi_{a,c}',
\end{align}
that describe the hybrid response of each resonator in the presence of two coherently summed driving terms.
They contain all the spectral information governing the Raman spectra. Indeed it can be shown that the pump enhancement of Eq.~\ref{eq:pump_enh} can also be written as a function of the effective susceptibilities when pumping only through one port (free space or waveguide). The final Raman spectrum will then be a product of the effective susceptibility squared magnitudes evaluated at the pump and Raman-shifted frequencies:
\begin{align}\label{eq:susc-product}
    S_{\text{ant,cav}}\propto \abs{\chi_{a,c}^\text{eff}(\omega_L)}^2 \abs{\chi_{a,c}^\text{eff}(\omega_D)}^2.
\end{align}
A fine tuning of the antenna-cavity detuning is then essential to maximize the Raman enhancement of the hybrid.


\section{Results}
\subsection{Hybrid SERS spectra}
\begin{figure*}
	\centering
	\includegraphics[width=0.9\linewidth]{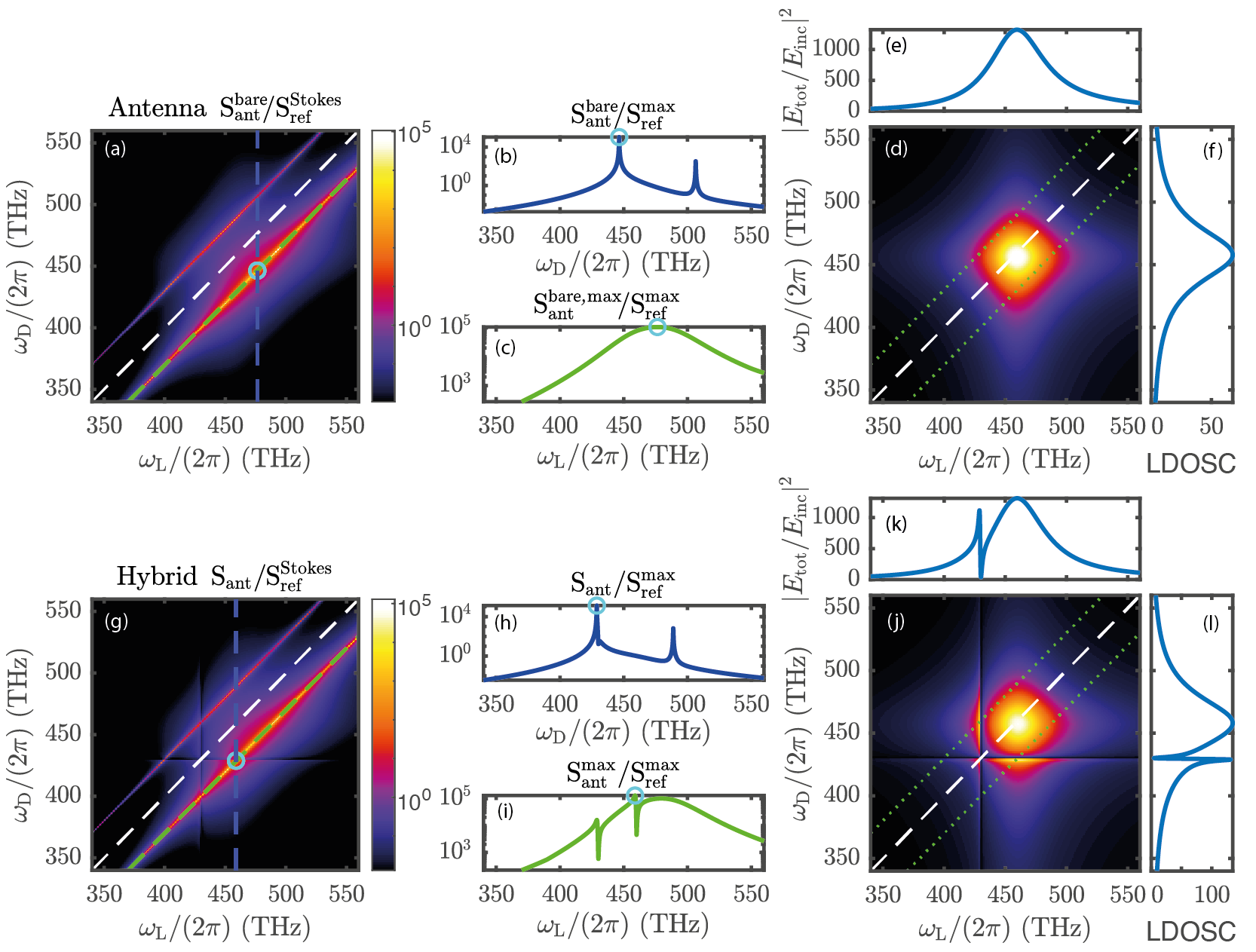}
	\caption{Raman spectrum enhanced by an antenna (a) as a function of the laser frequency, normalized by the Stokes emission peak of the molecules in air $S_{\mathrm{ref}}^{\mathrm{Stokes}}(\omega_L)$. The cross-cuts at the dashed blue and green line correspond respectively to (b) and (c). (b) Raman spectrum at the maximum enhancement, (c) antenna enhancement at the Stokes sideband as a function of the laser frequency. 
	(d) Raman spectrum of the bare antenna $S^{\text{bare}}_{\mathrm{ant}}$, normalized by the Raman emission of the molecules in air $S_{\mathrm{ref}}$ showing the antenna SERS enhancement as a function of the laser and detected frequencies. It is equal to the product of a pump enhancement term (e) and of a collected LDOS enhancement term (f). The case of a hybrid antenna-cavity resonator is given in (g-l), exhibiting narrow Fano resonances both in the Raman spectrum and in the pump and LDOS enhancements. See text for parameters.}
	\label{fig:single_mode_2d_and_cut}
\end{figure*}

In Fig.~\ref{fig:single_mode_2d_and_cut}(a) we plot the free-space Raman spectrum of an assumed Raman active species at a  bare antenna $S_\mathrm{ant}^\mathrm{bare}(\omega_L,\omega_D)$ , normalized by the Stokes peak amplitude  of the same analyte in a vacuum environment as reference $S_\mathrm{ref}^\mathrm{Stokes}=S_\mathrm{ref}(\omega_D=\omega_L-\Omega_m)$.
The antenna parameters are $\omega_a/(2\pi)=460\,\mathrm{THz}$, $\gamma_{i}/(2\pi)=20\,\mathrm{THz}$,  $\beta=0.12\,\mathrm{C^2.kg^{-1}}$ and an effective mode volume $V_a=3\left(\frac{\lambda}{10}\right)^3$ corresponding to the values of a dipole placed at 10 nm away from a $50\,\mathrm{nm}$ gold sphere. Free-space input and collection are assumed to occur via a NA=1 objective in vacuum, i.e. collection of the radiation in the upper half space. This results in $\eta_\mathrm{out}=1/2$ and excitation $\eta_\mathrm{in}=1/10$ due to finite scattering cross section (see Appendix~\ref{sec:Sup2}). The molecular vibration frequency in this example is chosen as $\Omega_m/(2\pi)=30\,\mathrm{THz}$, corresponding to typical Raman shifts of $1000\,\mathrm{cm^{-1}}$~\cite{LeRu2009PrinciplesSpectroscopy}, with quality factor $Q_m=200$.

The antenna-enhanced Raman spectrum shows two sidebands appearing as diagonals at $\omega_D=\omega_L\mp\Omega_m$, i.e. the anti-Stokes and Stokes sidebands observed at the laser frequency shifted by the mechanical vibration resonance frequency. A vertical cut at the maximum intensity (blue dashed line) is shown in Fig.~\ref{fig:single_mode_2d_and_cut}(b), representing a Raman spectrum at laser frequency fixed to the value at which the Stokes signal is most enhanced.  The detected Stokes signal when scanning the laser frequency (green diagonal dashed line, detection frequency shifting in concert with the laser frequency) is shown in Fig.~\ref{fig:single_mode_2d_and_cut}(c).  As in usual SERS experiments~\cite{LeRu2009PrinciplesSpectroscopy}, the maximum enhancement is achieved when the pump frequency is set at $\omega_L = \omega_a + \frac{\Omega_m}{2}$, resulting in the best trade-off between pump enhancement taking place at $\omega_L=\omega_a$, and emission enhancement happening at $\omega_D=\omega_a$. With the antenna and molecule position considered here, we obtain Raman enhancements on the order of $10^4$, limited only by the effective mode volume of the antenna.
The effect of the photonic system is better visualized in Fig.~\ref{fig:single_mode_2d_and_cut}(d) where we plot again the antenna enhanced Raman spectrum $S_\mathrm{ant}^\mathrm{bare}$, but now normalized to the reference Raman spectrum $S_\mathrm{ref}(\omega_D,\omega_L)$, to remove the dependence on the chosen mechanical vibration. We thus obtain the antenna enhancement compared to the homogeneous medium case for any pump and detection frequency. 
As it was derived for the hybrid case, the bare antenna Raman enhancement can be cast as a product of the pump enhancement $\abs{E_\text{tot}/E_\text{inc}}^2$ and the collected LDOS, shown respectively in Fig.~\ref{fig:single_mode_2d_and_cut}(e) and (f). Pump and LDOS enhancements for the bare antenna are obtained from Eq.~\ref{eq:pump_enh} and \ref{eq:LDOSac} by setting $J=G_{\text{cross}}=0$.
The pump enhancement depends on the laser frequency and the LDOSC on the detected frequency.
High enhancements of the Raman process are obtained by enhancing both the pump at $\omega_L$ and the LDOS at $\omega_L+\Omega_m$, and are thus usually achieved with broad antenna resonances such that $\gamma_a>\Omega_m$, placing them by default on the sideband non-resolved system.  

By a careful choice of parameters, the hybrid resonator allows to go beyond the limitation of sideband non-resolved optomechanics, yet obtain large SERS enhancement factors.
The same figures of Raman scattering spectra in the case of a hybrid antenna-dielectric resonator are shown in Fig~\ref{fig:single_mode_2d_and_cut}(g-l), with a cavity red detuned from the antenna by the mechanical vibration frequency $\omega_c/(2\pi)=(\omega_a-\Omega_m)/(2\pi)=430\,\mathrm{THz}$ corresponding to the Stokes sideband of the antenna, and a mode volume $V_c=10\lambda^3$ and quality factor $Q_c=10^3$. The main new feature is the appearance of a Fano resonance close to the cavity frequency, due to the interference of the coupled broad antenna resonance and the fine cavity resonance. This Fano feature is inherited both by the Raman spectrum, Fig~\ref{fig:single_mode_2d_and_cut}(h), and in the Stokes enhancement, Fig~\ref{fig:single_mode_2d_and_cut}(i), which show the capability of the hybrid system to obtain high enhancement with a high-Q resonance. 
Interestingly, the maximum Stokes enhancement, is obtained for a laser tuned at $\omega_L=\omega_a$, the hybrid resonator allowing to enhance the pump with the antenna resonance, and the emission with the cavity resonance. 
Better insight follows from the Raman enhancement of the hybrid compared to homogeneous medium and shown in Fig.~\ref{fig:single_mode_2d_and_cut}(j), which again is the product  of a pump enhancement, (Fig.~\ref{fig:single_mode_2d_and_cut}(k)) at the laser frequency, and a collected LDOS enhancement, (Fig.~\ref{fig:single_mode_2d_and_cut}(l)) at the detected frequency.
Both quantities display a  broad antenna-like resonance, and a narrow Fano resonance arising from mixing with the cavity-like mode.

\subsection{Choice of optimum read out scheme}
\begin{figure}
	\centering
	\includegraphics[width=1\linewidth]{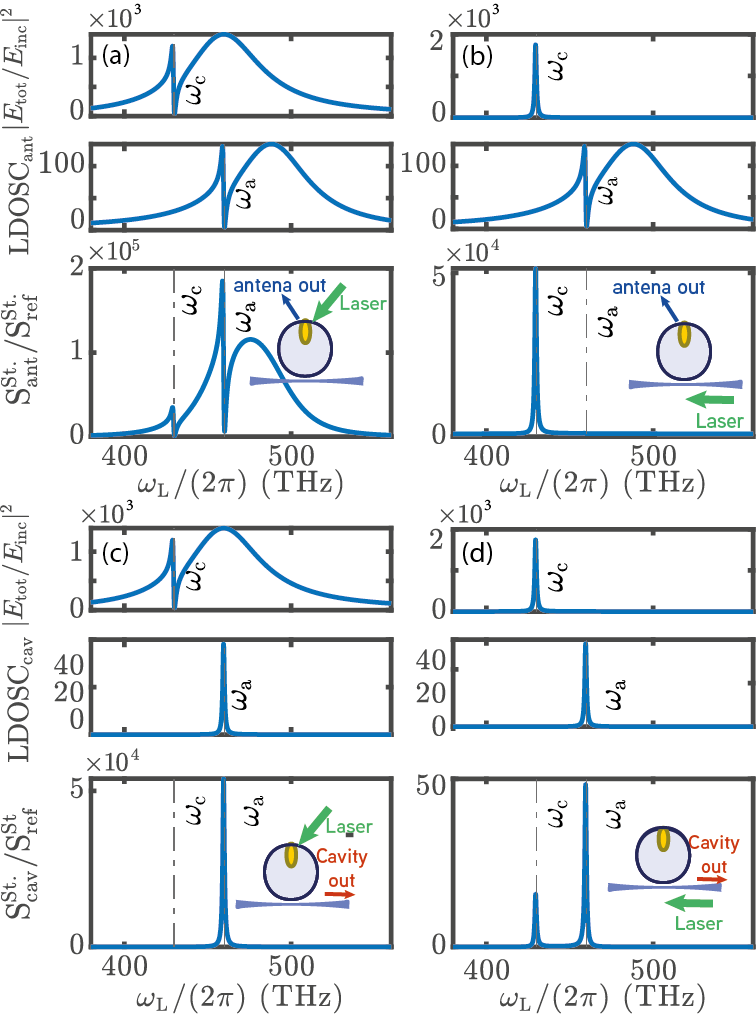}
	\caption{Stokes enhancement for the four different combinations of input output as depicted on the sketches. Each Stokes enhancement (iii) is the obtained as the product of the pump enhancement at $\omega_L$ (i) and the collected LDOS at $\omega_D=\omega_L-\Omega_m$ (ii) for the given input and output respectively. The parameters are the same as in Fig.~\ref{fig:single_mode_2d_and_cut}.}
	\label{fig:4inoutstokes}
\end{figure}

In the rest of the article we will focus on the Stokes (or anti-Stokes) enhancement curves, as the ones showed in Fig.~\ref{fig:single_mode_2d_and_cut}(c) and Fig.~\ref{fig:single_mode_2d_and_cut}(i), i.e. the detected frequency will be fixed at $\omega_D=\omega_L\mp\Omega_m$ as the laser frequency is scanned.
In the case of the hybrid, the two different input  and two different output ports  result in 4 different Raman spectroscopy scenarios depending on whether the pump and collection are performed through the waveguide or in free-space.
The Stokes enhancement factors for the four different cases are presented in Fig.~\ref{fig:4inoutstokes}.
The parameters are the same as in Fig.~\ref{fig:single_mode_2d_and_cut}, with again a cavity red detuned from the antenna by the molecule's vibration frequency, $\omega_c=\omega_a-\Omega_m$. 
The following observations can be made.  First, the simultaneous excitation and read-out through the waveguide acts as strong spectral filter at the cavity resonance.  Since pump and Raman signal are at shifted frequencies, this filtering action intrinsically results in low overall SERS enhancements, 3 orders of magnitude below that offered by just an antenna in free space.  Conversely, excitation and collection from free space results in large SERS enhancement, roughly on the same scale as the SERS enhancement that the bare antenna provides. However, the cavity mode elicits strong Fano features both when the pump and when the Stokes frequency go through cavity resonance.   Finally, we consider the `mixed port' cases where either the pump goes via the waveguide and collection is via free space, or vice versa.   Remarkably, the strongest pump field enhancement is reached when pumping through the waveguide and at cavity resonance $\omega_L \approx \omega_c$. For the chosen strongly blue detuned cavity the enhancement at the Stokes-shifted frequency for scattering in free space is modest due to the large detuning from antenna resonance, but nonetheless the joint effect is a strong SERS peak.  Conversely,  detection through the waveguide requires tuning $\omega_L = \omega _c + \Omega_m$. The pump field is resonantly enhanced by the antenna, while the Raman signal collection into the waveguide is enhanced over a narrow band around $\omega_D = \omega_c$.  The overall enhancement is similar to that in the reversed port choice to within a factor 2.
This result can appear surprising since only one of the two configurations is doubly resonant and one would expect better enhancement factors in this case. However the loss of enhancement due to a detuned antenna is compensated by the better output coupling efficiency for the antenna compared to the input coupling, $\eta_\text{in}=(1/5)\eta_\text{out}$ (see appendix~\ref{sec:Sup2}). The waveguide allows better incoupling efficiencies than the antenna but the collection efficiency can be, potentially, as good for the two.

To summarize, the hybrid system allows reaching Stokes enhancements of the same order of magnitude as the bare antenna case, but with much larger quality factors. Of particular interest for sideband resolved read out of vibrations is the case with collection through the waveguide where the Raman scattering is filtered by the narrow cavity resonance with enhancement factors similar to the case of free-space input and output. This allows to explore the sideband resolved regime with high Raman enhancement.
\subsection{Detuning dependence}
While coarsely speaking, the antenna and cavity frequencies need to be detuned by the mechanical vibration frequency $\omega_a-\omega_c=\pm\Omega_m$ to obtain the best enhancement factors, the fact that the enhancement is due to the product of Fano lineshapes imposes a finer analysis of the optimal detunings that are needed, and it is presented in the following. We focus here on the case where the cavity is red-detuned with respect to the antenna, and thus enhancing the Stokes sideband. Anti-stokes enhancement, requiring an inversed antenna-cavity detuning will be analyzed next.

\begin{figure}
	\centering
	\includegraphics[width=1\linewidth]{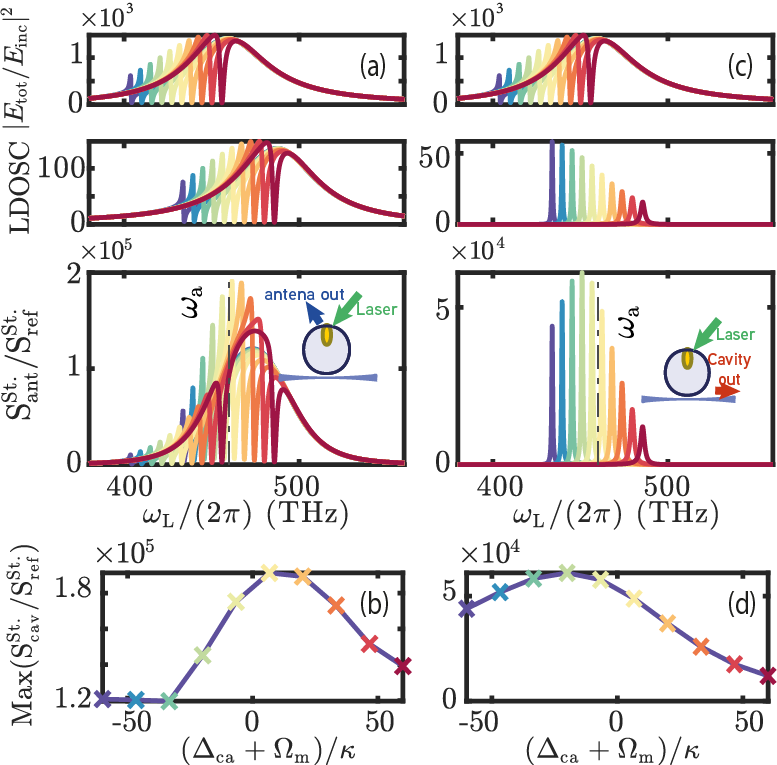}
	\caption{Influence of the cavity-antenna detuning $\Delta_{ca}$ on the Stokes enhancement for the free-space input/output (a) and free-space input and waveguide output (c). The antenna frequency is fixed at $\omega_a/(2\pi)=460\,\text{THz}$, and the cavity frequency is scanned around $\omega_a-\Omega_m$ in steps of $16\kappa$. The maximum Stokes enhancement for each detuning is shown in (b) and (d) for the two collection cases, with the colored crosses corresponding to the respective colored plot in (a) and (c). }
	\label{fig:best_input_output}
\end{figure}

Fig~\ref{fig:best_input_output} presents the Stokes enhancement factor for the two different collection ports, when pumping through free-space. The cavity-antenna detuning $\Delta_{ca}=\omega_c-\omega_a$ is changed by scanning the cavity frequency around $\omega_a-\Omega_m$, corresponding to the case considered in Fig~\ref{fig:single_mode_2d_and_cut}.
The cavity frequency is scanned by steps of $16\kappa$, with, for each frequency, the Stokes enhancement shown as the product of the pump enhancement and LDOSC. The pannels (b) and (d) show the maxima of the Stokes enhancement as a function of  detuning, each cross corresponding to the maximum Stokes enhancement of the same color in (a) and (c) respectively.
For both collection through free-space or in the waveguide the maximum achievable Stokes enhancement is obtained close to the intuitive detuning $\omega_c=\omega_a-\Omega_m$, but with some shift due to the Fano lineshapes. In the case of the waveguide output, an important contributor to the shift comes from an intrinsic asymmetry in antennas, namely the fact that radiative losses into free spaces decrease at low frequencies, which facilitates a higher overall coupling into the waveguide. 
Concerning the tuning sensitivity, since one of the two resonances in play is still the broad antenna resonance, the needed precision in the antenna-cavity detuning remains on the order of the antenna linewidth.

\subsection{Quality factor choice}\label{sec:QV}
\begin{figure}
	\centering
	\includegraphics[width=1\linewidth]{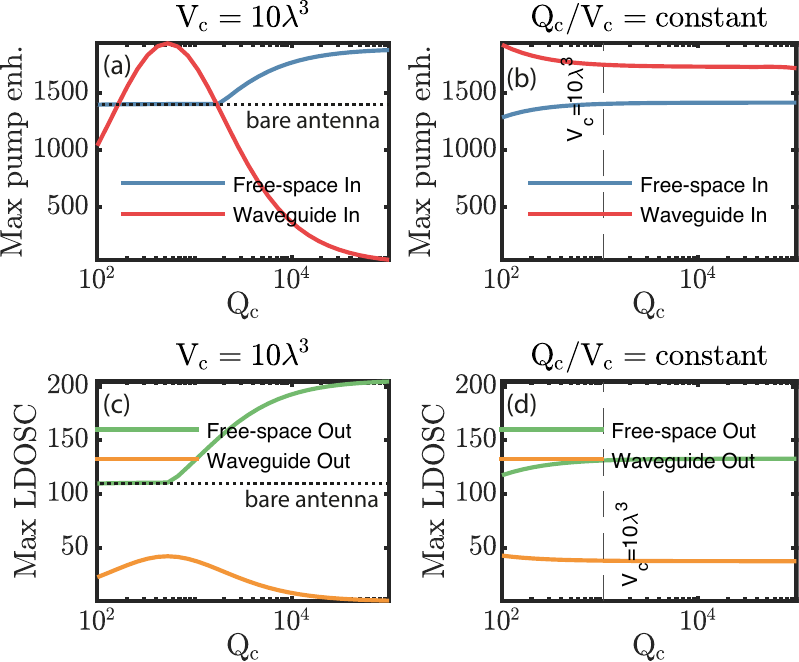}
	\caption{Maximum achievable pump enhancement (a,b) and LDOSC (c,d) as a function of the cavity quality factor $Q_c$ for both input and both output configurations. (a) and (c) are given for a fixed mode volume $V_c=10\lambda^3$, whereas (b) and (d) are given for a constant cavity Purcell factor $Q_c/V_c$. The antenna frequency is $\omega_a/(2\pi)=460\,\text{THz}$ and the cavity frequency is fixed at $\omega_c=\omega_a-\Omega_m$. Horizontal dotted line corresponds to the free-space case with only the bare antenna.}
	\label{fig:Qc_influence}
\end{figure}

An important question is how to choose the most appropriate cavity quality factor to reach the highest Raman enhancements. Aside from matching to the vibrational Q, the cavity Q will modify the in and outcoupling ratio into the waveguide compared to the free-space. This is analyzed in Fig.~\ref{fig:Qc_influence}, where we plot the maximum pump enhancement and collected LDOS of Eqs~\ref{eq:pump_enh}-\ref{eq:LDOSac} as a function of the cavity quality factor $Q_c$.
The antenna and cavity frequencies are $\omega_a/(2\pi)=460\,\text{THz}$, and $\omega_c=\omega_a-\omega_m$, corresponding to the double resonant case for simultaneous pump and collection enhancement.
Panels (a) and (c) first show  the case of a fixed mode volume $V_c=10\lambda^3$ (as used through-out this work). 
It can be seen that for free-space input and output, the enhancement factors are increased for higher $Q_c$, since the Fano resonances sharpen to  higher maximum values. Instead, for the case of waveguide input and ouput, there is an optimum at $Q_c\simeq1000$ both for the pump enhancement and LDOSC. This is due to a tradeoff with the cavity coupling efficiency that deterioriates for high $Q_c$ while for too small $Q_c$ the LDOS enhancement will be small. The exact value of the best $Q_c$ depends on the cavity mode volume $V_c$, which, determines the hybrid coupling efficiency through $\abs{J}^2\propto1/V_c$. 
For reference we also show the case of constant Purcell factor $Q_c/V_c$ cavity in Fig.~\ref{fig:Qc_influence}(b,d). It is seen that the LDOS and pump enhancement are mostly constant for the whole range of Purcell factors, showing that the LDOS of the hybrid resonances only depend on the ratio $Q_c/V_c$~\cite{Doeleman2016Antenna-CavityLinewidth}. This ratio is also proportionnal to $|J|^2/\kappa$ which appears in the denominator of Eq.~\ref{eq:suscept-hybrid} with $\chi_c(\omega_c)\propto Q_c$. This term dictates the hybrid interaction rate, and for constant $Q_c/V_c$, the interaction rate remains unchanged.

\subsection{Anti-stokes}
\begin{figure}
	\centering
	\includegraphics[width=1\linewidth]{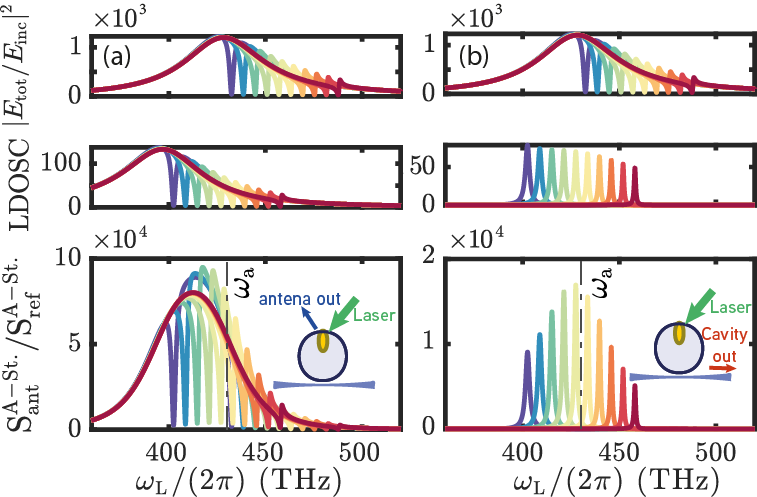}
	\caption{Anti-stokes enhancement for different cavity-antenna detunings. The antenna is now red detuned ($\omega_a=400\,\text{THz}$) to enhance the pump, and the cavity frequency is scanned around the anti-Stokes sideband ($\omega_a+\Omega_m$) in steps of $16\kappa$. The input is in free-space and the collection is either in free-space (a) or in the waveguide (b). }
	\label{fig:anti-stokes}
\end{figure}

Enhancement of the anti-Stokes sideband can also be achieved with the hybrid resonator. To achieve the best collection in the waveguide, the cavity now needs to be blue-detuned with respect to the antenna mode. As shown in  Fig.~\ref{fig:anti-stokes}, the enhancement is in this case slightly smaller than for the Stokes enhancement case. 
This is due to the increased radiative losses $\gamma_\text{rad}$ of the antenna at the (blue-shifted) collection frequency and a stronger emission from the reference dipole scaling as $\omega_D^4$ as seen in Eq.~\ref{eq:S_ref}.
This has an impact for both collection paths, which still results in a comparable enhancement factor for the waveguide collection case compared to the free-space collection.

\section{Multimode cavities}
\begin{figure}
	\centering
	\includegraphics[width=1\linewidth]{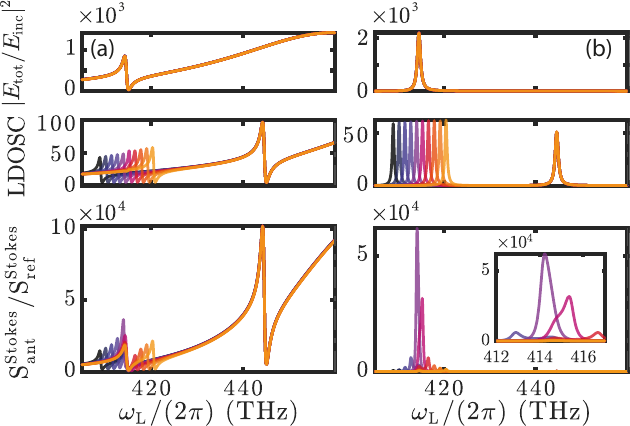}
	\caption{Stokes enhancement with a two-mode cavity and an antenna hybrid. The antenna is blue detuned $\omega_a/(2\pi)=460\,\text{THz}$ with respect to both cavity modes. The first cavity mode serves as pump enhancement at $\omega_P/(2\pi)=415\,\mathrm{THz}$ and the Stokes sideband emission is enhanced by the second cavity mode, which is scanned around $\omega_P-\Omega_m$. We compare the cases with free-space only (a) or waveguide only (b) input and output. The Stokes enhancement is again the product of a pump enhancement factor and the collected LDOS. Inset in (b) shows a closeup of the best Raman enhancements close to $\omega_L=\omega_P$.}
	\label{fig:multmode}
\end{figure}

We have shown that the hybrid with a single dielectric mode is able to provide integrated collection of the Raman signal with good enhancement factors for both Stokes and anti-Stokes. Nevertheless, a fully integrated operation is prevented by the sideband resolution of the cavity, that prevents from enhancing both the pump and the collection simultaneously through the waveguide. This issue can be resolved by working with multiple high-Q cavity modes, which allow both pump and collection enhancement. 
For instance, whispering gallery mode cavities provide multiple cavity modes addressable through  the same waveguide. In this way one could envision using two different cavity resonances to simultaneously enhance the  pump and collection, by tuning the mode spacing to match the mechanical resonance frequency.
The resulting Stokes enhancement factors are shown in Fig.~\ref{fig:multmode} for a hybrid with two cavity modes coupled to the same waveguide and an antenna coupled to free-space.
It can be shown that for multiple-resonant systems, the Raman enhancement can still be written as a product of the pump field enhancement by the hybrid, and the collection LDOS in a given port, similarly to Eq.~\ref{eq:factor_2}.
We compare the cases of only free-space addressing (a) with the fully integrated case (b). 
The first cavity mode is red detuned, to enhance the collection, and is labelled $C$. The second cavity will serve to enhance the pump and is labelled $P$. Both cavity modes are then red-detuned compared to the antenna so as to work in the optimal regime where the radiative losses of the antenna are reduced.
We have then from blue to red:
$\omega_a/(2\pi)=460\,\text{THz}$,
$\omega_P/(2\pi)=415\,\text{THz}$, and the frequency of the cavity mode use for collection scanned around the stokes sideband of the former  $\omega_C\simeq\omega_P-\Omega_m$ in steps of $1.1\kappa$.
We can see that the use of two cavity modes allows simultaneously a pump enhancement and an LDOS enhancement as shown in Fig.~\ref{fig:multmode}(b). This allows fully integrated Stokes enhancement factors that reach values equivalent to the Stokes enhancements of the bare antenna free-space configuration. This is furthermore obtained with high-Q resonances deep in the sideband resolved regime. Although the final enhancement -product of two hybridized cavity susceptibilities- necessitates fine frequency tuning on the order of $\kappa$, it allows for fully integrated Raman scattering with unspoiled enhancements.

\section{conclusion}
The analogies with cavity optomechanics have resulted in exciting predictions of new phenomena in the field of SERS. Most of these new applications, such as dynamical backaction~\cite{Roelli2016MolecularScattering} and low-noise THZ to optical transduction~\cite{Roelli2020}, benefit from both a good optomechanical coupling, and sideband resolution, i.e. an optical linewidth smaller than the Raman shift. This implies having a resonator that has both small mode volumes and high-Q resonance. 
Hybrid dielectric and plasmonic resonances can achieve simultaneously these two requirements, exploiting both the small volume of a plasmonic antenna, and the spectral confinement of dielectric cavities, with tunable parameters as a function of the detuning between the two resonators~\cite{Doeleman2016Antenna-CavityLinewidth}.

We have developed a new formalism based on molecular optomechanics, that allows to calculate absolute Raman enhancement factor of a multimode hybrid system from simple parameters of the bare constituents. The resulting expressions explicitly show the interplay of pump and LDOS enhancement factor. We have then demonstrated that using experimentally available~\cite{Doeleman2018ExperimentalContinuum} hybrid systems, one can reach Raman enhancement factors equivalent to the bare plasmonic case, but in the sideband resolved regime, with optical linewidths orders of magnitudes smaller than the mechanical frequency.
Additionally, our formalism correctly describes the coupling to different input and output ports, and we show that although optimal excitation and collection is reached through the antenna port, Raman enhancements for collection in the waveguide remains on the same order of magnitude.
Finally, an efficient and fully integrated platform is proposed using simultaneously two different cavity modes, hybridized with a plasmonic antenna and coupled to same waveguide, each enhancing the pump and the collection respectively. 
Although we have here focused on the peak enhancement factors, the opportunities for Fano lineshapes are particularly exciting in quantum optomechanical applications employing reservoir engineering~\cite{Yanay2018ReservoirDissipation}.

\appendix
\section{Derivation of the effective hybrid Hamiltonian}\label{sec:Sup1}
The optomechanical coupling is described as a shift of the antenna resonance frequency due to the mechanical motion. Introducing the position operator, $\hat{x}_m=x_\text{zpf}(\hat{b}^{\dagger}+\hat{b})$, the frequency shift of an optical mode can be expressed to first order in $x_m$ as~\cite{Roelli2016MolecularScattering}:
\begin{equation}
    \omega_a(x_m)=\omega_a-G_m {x}_m,
\end{equation}
with the optomechanical coupling rate $G_a\equiv-\frac{\partial \omega_a}{\partial x_m}$. 
Crossed optomechanical interaction appears when multiple optical modes interact with the same mechanical resonator~\cite{Cheung2011,Biancofiore2011}. The crossed optomechanical rate is proportional to the overlap of the optical fields at the surface of the mechanical resonator. In the case of molecular optomechanics where the mechanical resonator is considered to be a point dipole, the crossed optomechanical coupling simplifies to $G_\text{cross}=\sqrt{G_a G_c}$. Alternatively one can use a dipolar interaction Hamiltonian between the molecule's Raman dipole $\mathbf{\hat{p}}_R$ and the optical fields at the molecules position $\mathbf{r_m}$~\cite{Schmidt2017LinkingSERS},
\begin{equation}
    H_\text{I}=-\frac{1}{2}\mathbf{\hat{p}}_R(t)\cdot \mathbf{\hat{E}}(\mathbf{r}_m,t).
\end{equation} The total field of the antenna and the cavity at molecule's position, along the dipole of the molecule is \begin{equation}
{\hat{E}}(\mathbf{r}_m,t)=\sqrt{\frac{\hbar\omega_c}{2V_c\epsilon_0}}(\hat{a}(t)+\hat{a}^\dagger(t))+
\sqrt{\frac{\hbar\omega_a}{2V_a\epsilon_0}}(\hat{c}(t)+\hat{c}^\dagger(t)).\label{eq:fieldsup}
\end{equation}
The Raman dipole operator can be written as a function of the Raman tensor as   $\hat{p}_R(t)=\frac{\partial \alpha}{\partial x_m}x_m\hat{E}(\mathbf{r}_m,t)$.
Inserting this expression and the electric field of Eq.~\ref{eq:fieldsup} into the interaction Hamiltonian, and discarding non-resonant terms yields the optomechanical interaction Hamiltonian, with the crossed optomechanical coupling.
Next, the hybrid coupling between optical modes is obtained through a Green tensor approach~\cite{Doeleman2016Antenna-CavityLinewidth}, which considers the antenna as a single polarizable dipole. This model can directly be mapped on a quantum optics formalism~\cite{Medina2021}.
The total Hamiltonian is finally written ($\hbar=1$):
\begin{align}
    \hat{H}&=\omega_a\hat{a}^{\dagger}\hat{a}+\omega_c\hat{c}^{\dagger}\hat{c}+\Omega_m\hat{b}^{\dagger}\hat{b}\\ \nonumber&-x_{\text{zpf}}\left( G_a \hat{a}^{\dagger}\hat{a}+ G_c \hat{c}^{\dagger}\hat{c}+ G_\text{cross} (\hat{a}^{\dagger}\hat{c} +\hat{c}^{\dagger}\hat{a})
    \right)(\hat{b}^{\dagger}+\hat{b})\\ \nonumber&
    +J(\hat{a}^{\dagger}\hat{c} +\hat{c}^{\dagger}\hat{a})+\hat{H}_{dr}.
    \label{eq:Hamiltonian}
\end{align}
The annihilation operators for the antenna and cavity modes are $\hat{a}$ and $\hat{c}$, and the driving of the antenna or the cavity by a laser is described by $\hat{H}_{dr}\propto a+a^\dagger$ or $\propto c+c^\dagger$. The mechanical displacement operator is $\hat{x}=x_{\mathrm{zpf}}(\hat{b}+\hat{b}^\dagger)$, with $x_{\mathrm{zpf}}$ the mechanical zero-point fluctuation. The classical Langevin equations describe the evolution of the expectation values of these three operators. We also consider the high photon number (mean-field) limit where $\left<\hat{x}_m\hat{a}\right>=\left<\hat{x}_m\right>\left<\hat{a}\right>$~\cite{bowen_2015}. The resulting classical Langevin equations are given in~\ref{eq:langevin}.

\section{Input and ouput parameters}\label{sec:Sup2}
The source terms for the laser pump have been written such that $\abs{s_{\text{in},a}}^2$ and $\abs{s_{\text{in},c}}^2$ correspond to the optical power arriving through the free-space and the waveguide. 
The input coupling efficiencies dictate the portion of the input power that is effectively coupled into each resonator, and are written as a fraction of the total decay rate of each resonator.
For the waveguide, the input and output coupling are chosen to be $\eta_{\text{in,c}}=\eta_{\text{out,c}}=1/4$ (critical coupling). 
For the antenna we have assumed a diffraction limited focusing of a collimated input beam, from which we can write the incoming photon flux as 
\begin{align}
     \abs{s_{a,\text{in}}}^2&=\pi\left(1.22\frac{\lambda}{2}\right)^2\frac{\epsilon_0 c}{2}\abs{E_\text{inc}}^2,
\end{align}
with $E_\text{inc}$ the incoming electric field.
By using the equation of motion for the antenna field as a function of the antenna dipole moment $p$ in the rotating wave approximation obtained from a Green-function based analysis~\cite{Doeleman2016Antenna-CavityLinewidth}:
\begin{align}
    \left(\omega_a-\omega-i\frac{\gamma}{2}\right)p-\frac{\beta}{2\omega}\tilde{E}_c c = \frac{\beta}{2\omega} E_{\text{inc}}
\end{align}
and comparing it to Eq.~\ref{eq:langevin} one obtains:
\begin{align}
    \sqrt{\eta_{a,\text{in}}\gamma_\text{rad}}s_{a,\text{in}} \equiv -i\sqrt{\frac{\beta}{8}}E_{\text{inc}},
\end{align}
where we have used $\gamma_\text{rad}=\frac{\beta\omega^2}{6\pi\epsilon_0c^3}$ and $p=\frac{\sqrt{2\beta}}{\omega}a$.
From this one can express the input coupling efficiency for the antenna
\begin{align}
 \eta_{a,\text{in}}&=\frac{\beta}{8\gamma_\text{rad}}\frac{1}{\pi\left(0.66\lambda\right)^2\frac{\epsilon_0 c}{2}}\nonumber\\
     &=\frac{27}{32\pi^2}
\end{align}
which we have approximated to $\eta_{a,\text{in}}=1/10$ throughout the article.
Collection of the emission in the upper half-space yields a collection efficiency of $\eta_{a,\text{out}}=1/2$. Thus, due to reduced exctinction cross section of a dipolar scatterer, the input and output coupling efficiencies are not equal, and collection efficiency is roughly 5 times more efficient than excitation efficiency.
It should be noted that it is possible for the input and output efficiencies to be different as the free-space radiation channel is in fact composed of a continuum of modes, and the input field and radiation (output) fields are not distributed over those modes equally.

\begin{acknowledgments}
This work is part of the research program of the Netherlands Organisation for Scientific Research (NWO). The authors acknowledge support from the European Union’s Horizon 2020 research and innovation program under Grant Agreements No. 829067 (FET Open THOR) and No. 732894 (FET Proactive HOT), and the European Research Council (ERC starting Grant No. 759644-TOPP). The authors thank Javier del Pino and Philippe Lalanne for fruitful discussions and for their support.
\end{acknowledgments}

\bibliography{biblio}
\end{document}